\documentclass[floats,preprint,superscriptaddress,tightenlines]{revtex4}
\usepackage{bm}
\usepackage{epsfig}
\usepackage{rotating}

\newcommand{\fr}{\frac}
\newcommand{\ra}{\rightarrow}
\newcommand{\ov}{\overline}

\newcommand{\nn}{\nonumber}
\newcommand{\beq}{\begin{equation}}
\newcommand{\eeq}{\end{equation}}
\newcommand{\bea}{\begin{eqnarray}}
\newcommand{\eea}{\end{eqnarray}}
\newcommand{\ben}{\begin{eqnarray*}}
\newcommand{\een}{\end{eqnarray*}}

\def\vec#1{{\bf #1}}
\def\D0{D\O}

\newcommand{\tenb}{\bf{\overline{10}}}
\begin{document}
\preprint{ \vbox{\hbox{DUKE-TH-04-260}}} 

\title{Determining Pentaquark Quantum Numbers from Strong Decays}%

\author{Thomas Mehen}%
\email{mehen@phy.duke.edu}
\affiliation{Department of Physics, Duke University, Durham NC 27708, USA}
\affiliation{Jefferson Laboratory, 12000 Jefferson Ave., Newport News VA 23606}

\author{Carlos Schat}%
\email{schat@phy.duke.edu}
\affiliation{Department of Physics, Duke University, Durham NC 27708, USA}

\begin{abstract}
Assuming that the recently observed $\Theta^+$  and $\Xi^{--}$
are members of an anti-decuplet of $SU(3)$, decays to ground state
baryons and mesons are calculated using an effective  Lagrangian 
which incorporates chiral and $SU(3)$ symmetry. We consider the possible quantum 
number assignments $J^\Pi =\frac{1}{2}^\pm,\frac{3}{2}^\pm$
and calculate  ratios of partial widths. The branching ratios of exotic cascades  can be used to 
discriminate between even and odd parity pentaquarks. 
\end{abstract} 

\date{\today}
\maketitle

The newly discovered $\Theta^+(1540)$~\cite{Nakano:2003qx,Stepanyan:2003qr,Kubarovsky:2003fi,
Barmin:2003vv,Barth:2003es,Asratyan:2003cb,Airapetian:2003ri} is the first known example of a manifestly exotic hadron.
More recently, the NA49 collaboration has seen an exotic $\Xi^{--}$ with a mass of approximately 1860 MeV as
well as evidence for a $\Xi^0$ with nearly the same mass~\cite{Alt:2003vb}.  The $\Theta^+$ has baryon number
$B=1$ and strangeness $S=1$ and is interpreted as a $(uudd\bar s)$  pentaquark while the $\Xi^{--}$ has $B=1$
and $S=-2$ and is interpreted as a $(ddss\bar u)$ pentaquark. The bounds on the  widths of the $\Theta^+$
and $\Xi$  are quite small for strongly decaying hadronic resonances. These 
hadrons provide a new arena in which to test and improve our understanding of low energy QCD.
Much current experimental and theoretical effort is focused on understanding their properties.

The most pressing experimental problems are searching for the $SU(3)$ partners of the $\Theta^+$ and $\Xi^{--}$ and
determining the quantum numbers of these particles. While there is some evidence that the $\Theta^+$ and $\Xi^{--}$ are
members of the $\tenb$  representation of $SU(3)$, at the present time there are no published experimental  constraints on
the spin and parity of these states. Some proposals exist for measuring the parity of the $\Theta^+$ in polarized
proton-proton collisions  \cite{Thomas:2003ak,Hanhart:2003xp} and in photoproduction~\cite{photomodels,Rekalo:2004it}. 
In this letter, we
use chiral perturbation theory to analyze the decays $P \to B M$,  where $P$ is a pentaquark in the $\tenb$ of $SU(3)$, and
$B$ and $M$ are members of the ground state octet baryons and mesons, respectively.   On general grounds, two-body decay
rates to ground state nucleons and Goldstone bosons scale as 
\bea\label{sup}
\begin{array}{cc} p^{2 L+1} & (L>0) \, ,  \\
E^2 p & (L=0) \, ,
\end{array} 
\eea  
where $E$ is the energy of the Goldstone boson and $p = |\vec p|$ is its momentum.
($S$-wave decays scale as $E^2 p$ rather than $p$ because the Goldstone bosons are derivatively coupled.) This information
can be combined with $SU(3)$ symmetry  to make model-independent, parameter-free predictions for  ratios of partial widths
which are sensitive to the partial wave of the decay and can be used  to constrain the possible quantum numbers of the
pentaquarks. Pentaquarks with  $J^\Pi =\frac{1}{2}^-$ decay via $S$-waves, $J^\Pi=\frac{1}{2}^+,\frac{3}{2}^+$ states decay
via $P$-waves, $J^\Pi=\frac{3}{2}^-,\frac{5}{2}^-$ via $D$ waves, etc. Therefore, while it is impossible to distinguish
$\frac{1}{2}^+$ from $\frac{3}{2}^+$ or  $\frac{3}{2}^-$ from $\frac{5}{2}^-$ using these ratios, discrimination between
even and odd parity states is possible.  Of course, it is possible to measure the partial wave from the angular
distribution of  the meson in the decay. However, this requires knowledge of the polarization of either the initial
pentaquark or final state baryon as well as greater statistics to reconstruct an angular distribution. We focus on the
partial widths rather than angular distributions because they may be more accessible experimentally.

Calculations of pentaquark decays within chiral soliton models~\cite{Diakonov:1997mm}  and constituent quark 
models~\cite{Carlson:2003wc,Carlson:2003pn} already exist. These analyses predate the discovery of the $\Xi^{--}$ and
therefore use model predictions for the $\Xi^{--}$ mass which are no longer appropriate. Furthermore, they only consider
either $J^\Pi= \frac{1}{2}^+$ or $\frac{1}{2}^-$. In this  paper, we calculate the decays within the  model-independent
framework of heavy baryon chiral perturbation theory~\cite{Jenkins:1990jv} using experimentally measured masses. We
consider the decays of $J^\Pi=\frac{3}{2}^\pm$ as well as $J^\Pi = \frac{1}{2}^\pm$ pentaquarks and estimate uncertainties
due to $SU(3)$ breaking. A few of our results overlap with a recent analysis of $\Xi$ decays that focuses mostly on
pentaquarks with $J^\Pi = \frac{1}{2}^+$~\cite{Jaffe:2003ci}. We show that  measurement of the ratios $\Gamma[\Xi^{--} \to
\Xi^- \pi^-]/\Gamma[\Xi^{--} \to \Sigma^- K^-]$ and $\Gamma[\Xi^{+} \to \Xi^0 \pi^+]/\Gamma[\Xi^{+} \to \Sigma^+
\overline{K}^0]$ can reliably distinguish between $S$ and $P$-wave decays of the pentaquark. Since $\Theta^+$ only decays
to $N K$ the total width can be determined  from two-body decays. The $\Xi$ states are heavy enough to have other decay
modes so our calculations of two-body decays only give a lower limit on the ratio $\Gamma[\Xi]/\Gamma[\Theta^+]$. Measuring
$\Gamma[\Xi^{--}], \Gamma[\Xi^{+}]< 10 \,\Gamma[\Theta^+]$ can rule out $J^\Pi =\frac{3}{2}^-$ and  $J \geq \frac{5}{2}$.
We also predict $\Gamma[\Xi^{--}], \Gamma[\Xi^{+}] \geq 4 \, \Gamma[\Theta^+]$ if  $J^P=\frac{1}{2}^+ , \frac{3}{2}^+$
while the lower bounds are smaller for $J^\Pi=\frac{1}{2}^-$. 

The smallest $SU(3)$ representation that can accommodate the $\Theta^+$ is the $\tenb$. Larger representations 
require isospin partners of the $\Theta^+$ which have not been 
observed~\cite{Barth:2003es,Juengst:2003yy,Airapetian:2003ri,Capstick:2003iq}. 
There are hints  that the  $\Xi^{--}$ is also a
member of the $\tenb$. The $\Xi^{--}$ has $I_3 = -\frac{3}{2}$ and therefore must have an  $I_3 = \frac{3}{2}$
partner, $\Xi^+$. Preliminary results from Ref.~\cite{kadija} show evidence for the decay $\Xi^-  \to
\Xi^0(1530)\pi^-$   but not the decay $\Xi^+ \to \Xi^0(1530)\pi^+$, where the well established 
$\Xi^0(1530)$ resonance belongs to a ${\bf{10}}$ and has  $J^\Pi =\frac{3}{2}^+$.  The decays  $\Xi^\pm  \to
\Xi^0(1530)\pi^\pm$ are not allowed if the $\Xi^\pm$ are members of the $\tenb$ representation of $SU(3)$ but
are allowed for  the $\bf 8$, which contains a $\Xi^-$ but not a $\Xi^+$, and larger representations such as
the ${\bf 27}$ or ${\bf 35}$, which contain both $\Xi^+$ and $\Xi^-$. As pointed out  in
Ref.~\cite{Jaffe:2003ci} the observation of $\Xi^-  \to \Xi^0(1530)\pi^-$ but not $\Xi^+  \to
\Xi^0(1530)\pi^+$ can be understood if the $\Xi^{+}$, and hence  $\Xi^{--}$, are members of the $\tenb$
and the $\Xi^-$ reported in Ref.~\cite{kadija} belongs to a different $SU(3)$ multiplet.  For the
remainder of this paper, we will assume that the $\Theta^+$ and $\Xi^{--}$ are members of a $\tenb$  multiplet.

Other quantum numbers of the pentaquark such as the spin and parity  are  unconstrained by present data. A number  of
different theoretical interpretations of the pentaquark exist which  give different predictions for these quantum numbers.  
The chiral soliton model of Ref.~\cite{Diakonov:1997mm} predicted  a narrow pentaquark with a mass of 1530 GeV with
$J^\Pi=\frac{1}{2}^+$ prior to the discovery of the $\Theta^+$. The same quantum numbers arise in the correlated diquark
model in which the four quarks of the pentaquark are first paired into diquarks and  then placed in an $L=1$ partial
wave~\cite{Jaffe:2003sg}. The conventional  quark model can give rise to  both $J^\Pi= \frac{1}{2}^+$ or
$J^\Pi=\frac{1}{2}^-$ pentaquarks, depending on whether or not the quarks in the ground state have orbital angular 
momentum~\cite{Carlson:2003xb,Carlson:2003wc, Carlson:2003pn,Karliner:2003sy,Stancu:2003if,Jennings:2003wz,Glozman:2003sy}.
Other calculations using QCD sum rules~\cite{Zhu:2003ba} or the lattice~\cite{Csikor:2003ng,Sasaki:2003gi} 
favor negative parity assignments for the lowest mass pentaquark state. Pentaquarks with $J^\Pi=\frac{3}{2}^+$
have been considered in Ref.~\cite{threehalf}. All hadronic models predict either that $J^\Pi=\frac{1}{2}^+$ or
$J^\Pi=\frac{1}{2}^-$ is the lowest mass pentaquark.

Some recent papers have been devoted to the problem of determining  the quantum numbers of the $\Theta^+$
and $\Xi^{--}$ experimentally. An interesting idea suggested in Ref.~\cite{Thomas:2003ak}  is to study $\Theta^+$
production near threshold in polarized proton-proton collisions. At threshold a positive (negative) parity
$\Theta^+$ can be produced only by  $pp$ in a spin singlet (triplet) state. This observation relies only  on angular
momentum and parity conservation and is therefore  independent of any dynamical model of the production
mechanism. The idea was further developed  in Ref.~\cite{Hanhart:2003xp}, where  spin asymmetries at and
slightly above threshold were calculated in a model independent fashion.  Other proposals for extracting the
parity from spin asymmetries  in photoproduction of $\Theta^+$ were discussed in Ref.~\cite{photomodels}.  The
calculations of these papers rely on field theoretic models of the $\Theta^+$  production process. Because the
energy of the photon in the lab frame at the $\Theta^+$ threshold is $1.8$ GeV,  the process is beyond
the range of applicability of chiral perturbation theory, so a systematic calculation with controlled errors
is not possible for this process. A model independent analysis of polarized photoproduction 
recently appeared in Ref.~\cite{Rekalo:2004it}.

Pentaquark decays are a more promising application of chiral perturbation theory. In the   decays,
$\Theta^+ \to n K^+$, $\Xi^{--} \to \Xi^- \pi^-$ and  $ \Xi^{--} \to \Sigma^- K^-$, the momenta of the
Goldstone bosons  are 270 MeV, 445 MeV and 360 MeV, respectively. $SU(3)$ chiral perturbation theory is an
expansion in $Q/\Lambda_\chi$ where $\Lambda_\chi \sim 4 \pi f_\pi \approx 1.2$ GeV and $Q \sim m_\pi, m_K \sim p$,
where $p$ is a typical momentum for the process under consideration. Therefore, pentaquark decays can be
analyzed perturbatively using chiral perturbation theory.  We will analyze the decays of $J^\Pi=
\frac{1}{2}^{\pm}, \frac{3}{2}^\pm$ pentaquarks using the heavy baryon chiral perturbation theory formalism of
Ref.~\cite{Jenkins:1990jv}. The $\tenb$ pentaquarks  are contained in the completely symmetric tensor field
$P_{ijk}$:
\bea
P_{333} &=& \Theta^{+}  \\
P_{133} = \frac{1}{\sqrt{3}} \, N^0_{\tenb}  & & 
P_{233} = \frac{1}{\sqrt{3}} \, N^+_{\tenb} \nn \\
P_{113} = \frac{1}{\sqrt{3}} \, \Sigma^{-}_{\tenb}\qquad 
P_{123} &=& \frac{1}{\sqrt{6}}\Sigma^{0}_{\tenb}\qquad 
P_{223} = \frac{1}{\sqrt{3}} \, \Sigma^{+}_{\tenb}   \nn \\
P_{111} =  \Xi^{--}_{\tenb} \qquad 
P_{112} = \frac{1}{\sqrt{3}}  \, \Xi_{\tenb}^{-}  & & 
P_{122} = \frac{1}{\sqrt{3}} \, \Xi_{\tenb}^{0} \qquad 
P_{222} =  \Xi_{\tenb}^{+}\, . \nn
\eea
The subscript $\tenb$ on the $N, \Sigma$, and $\Xi$ fields serves
to distinguish these states from analogous states  in the ${\bf 8}$ of $SU(3)$.  An octet of $SU(3)$ pentaquarks
nearly degenerate with the $\tenb$ is predicted by the diquark model
of Ref.~\cite{Jaffe:2003sg}.

There is a unique $SU(3)$ invariant coupling of
the baryon octet, $B^{i}_{\; l}$, the Goldstone boson octet,
$M^j_{\; n}$, and the pentaquark $P_{ijk}$~\cite{Oh:2003fs}:
\bea \label{clebsch}
\overline{B}^{i}_{\; l} M^j_{\; n} P_{ijk}\epsilon^{l n k} &=& \fr{1}{\sqrt{2}} \ov K^0 \ov p \Theta^+ 
- \fr{1}{\sqrt{2}} K^- \ov n \Theta^+ 
- \fr{1}{\sqrt{2}} \pi^+  \ov \Xi^-\Xi^{--}_{\tenb} 
+  \fr{1}{\sqrt{2}} K^+  \ov \Sigma^- \Xi^{--}_{\tenb}  \\  
& & + \fr{1}{\sqrt{6}} \pi^- \ov \Xi^- \Xi^0_{\tenb}
+ \fr{1}{\sqrt{3}} \pi^0 \ov \Xi^0 \Xi^0_{\tenb}
- \fr{1}{\sqrt{6}} K^+ \ov \Sigma^+ \Xi^0_{\tenb}
- \fr{1}{\sqrt{3}} K^0 \ov \Sigma^0 \Xi^0_{\tenb}  + ... \, .  \nn
\eea 
In our expressions for the  Lagrangian the $SU(3)$ indices will be suppressed.
The kinetic terms in the heavy baryon chiral Lagrangian are
\bea\label{kin}
{\cal L}_{kin} &=& i \overline B ( v \cdot D ) B  + \fr{f^2}{4} \partial_\mu \Sigma \partial^\mu \Sigma^\dagger 
+ {\cal L}^{\frac{1}{2}^\pm , \, \frac{3}{2}^\pm }_{kin} \, , \\
{\cal L}^{\frac{1}{2}^\pm}_{kin} &=&   \overline P_\pm (i v \cdot D - \Delta) P_\pm \; , \qquad  \qquad
{\cal L}^{\frac{3}{2}^\pm}_{kin} =  \overline P^{\mu}_\pm ( - i v \cdot D  + \Delta) P_{\mu}^\pm \nn 
\; ,
\eea 
where $\Delta = m_P - m_B \sim 500$ MeV is the residual mass term, the $J^\Pi=\frac{1}{2}^\pm$ pentaquark field is
$P_\pm$,  the $J^\Pi=\frac{3}{2}^\pm$ pentaquark field is $P_\pm^\mu$,   $D_\mu$ is the chiral covariant derivative,  $
\Sigma = \xi^2 = \exp(2 i M / f) $  with $f = f_\pi \approx 93 \, \rm MeV$.    Since $m_B \sim m_P \sim \Lambda_\chi$, it
is necessary to perform a field redefinition on the baryon fields to obtain manifest power counting in the
Lagrangian~\cite{Jenkins:1990jv,Jenkins:1991ne}. The baryon and pentaquark fields then have static propagators and
derivatives acting on the $B$ or $P$ fields bring factors of the residual momentum which is  $O(Q)$. The residual mass is
approximately the same size as $m_K$, so it is also $O(Q)$.  A  relativistic version of chiral  perturbation theory for
pentaquark interactions is introduced in Ref.~\cite{Ko:2003xx}. This paper gives expressions for two-body decay 
widths of $J^\Pi =\frac{1}{2}^\pm$ pentaquarks that are used to normalize calculations of 
production cross sections. If the  relativistic expressions are used  
to make  predictions for the partial width ratios considered below,  the results differ from  the
values quoted in the present paper by only 10-15\%. This is smaller than the size of corrections expected from $SU(3)$ breaking.

Next we write down the leading interaction Lagrangians for each of the 
cases $J^\Pi = \frac{1}{2}^\pm, \frac{3}{2}^\pm$. For $J^\Pi=\frac{3}{2}^-$,
the leading interaction is $O(Q^2)$, for all others the leading interaction 
is $O(Q)$. In each case there is a unique term.
For a $J^\Pi=\frac{1}{2}^+$ pentaquark, the interaction Lagrangian is
\bea
{\cal L}^{\frac{1}{2}^+} = 2 g \, (\overline B \, S^\mu \, A_\mu \, P_+  + h.c.) \, ,
\eea
where $A_\mu = \frac{i}{2}(\xi \partial_\mu \xi^\dagger  - \xi^\dagger \partial_\mu\xi) 
= \partial_\mu M/f + ...$, and $S_\mu$ are the spin operators introduced in Ref.~\cite{Jenkins:1990jv}.
The interaction Lagrangian for a $J^\Pi=\frac{1}{2}^-$ pentaquark is
\bea
{\cal L}^{\frac{1}{2}^-} = g \,(\overline B \,v \cdot A \, P_- + h.c.) \, ,
\eea
the interaction Lagrangian for $J^\Pi=\frac{3}{2}^+$ is 
\bea
{\cal L}^{\frac{3}{2}^+} = g \,(\overline B  \,A_\mu \,P_+^\mu  + h.c.) \, ,
\eea 
and the interaction Lagrangian for  $J^\Pi=\frac{3}{2}^-$ is
\bea\label{thminus}
{\cal L}^{\frac{3}{2}^-} = \frac{g}{\Lambda_\chi} \left[\overline B (i D_\mu A_\nu 
+i D_\nu A_\mu) 
S^\nu P_-^\mu + h.c.\right]   \, .
\eea
The dimensionless coupling constants in each of the interactions are unrelated. Since only one is relevant 
for a given $J^\Pi$, we will denote all of them by the same symbol $g$.

The decay rates are given by
\bea
\Gamma(P \to B M) = ({\cal C.G.})^2\frac{g^2}{2 \pi f^2} \frac{m_B}{m_P} 
\left\{ \begin{array}{cc} 
 E^2 p  & J^\Pi = \frac{1}{2}^-\\ 
 p^3    & J^\Pi = \frac{1}{2}^+ \\
\fr{1}{3}  p^3    & J^\Pi = \frac{3}{2}^+ \\
\frac{1}{3 \Lambda_\chi^2} p^5 & J^\Pi = \frac{3}{2}^- 
\end{array} \right. \,.
\eea
The factor ${\cal C.G.}$ is an $SU(3)$ Clebsch-Gordan coefficient that can be read off 
by expanding  $\overline{B}^{i}_{\; l} M^j_{\; n} P_{ijk}\epsilon^{l n k}$  in component
fields as shown in Eq.~(\ref{clebsch}). For the $\Theta^+$ total width we obtain:  
\bea\label{rates}
\Gamma(\Theta^+) = \Gamma(\Theta^+ \to n K^+) + \Gamma(\Theta^+ \to p K^0) =
g^2 \left\{ \begin{array}{cc} 
646 \,{\rm MeV}  & J^\Pi = \frac{1}{2}^-\\ 
146 \,{\rm MeV}  & J^\Pi = \frac{1}{2}^+ \\
49 \,{\rm MeV}  & J^\Pi = \frac{3}{2}^+ \\
2.4 \, {\rm MeV}  & J^\Pi = \frac{3}{2}^- 
\end{array} \right. \, .
\eea
In obtaining these results we have made the replacement $f \to f_K = 1.22 f_\pi$ since these decays
involve a kaon. The procedure of using $f_K$ for decays involving kaons and $f_\pi$ for decays with pions
is a common practice and we will do this for all calculations in this paper. This procedure
incorporates some known $SU(3)$ breaking effects coming from the renormalization of the meson decay constants.
Since  this is not a systematic calculation of higher order effects, uncertainties in our predictions
are still about $\sim 30$ \%,  which is the typical size of violations of  $SU(3)$ symmetry.
 
Bounds from photoproduction  experiments are
limited by the detector resolution and give  $\Gamma[\Theta^+] < 20-25$ MeV while the DIANA $K^+ Xe$ experiment places a
tighter bound of $\Gamma[\Theta^+] < 9$ MeV~\cite{Barmin:2003vv}.  Indirect bounds coming from the analysis of older $K d$
and $K N$ scattering data as well as reanalysis of  $K^+ Xe$ data claim bounds on widths as low as $\approx  1$ MeV \cite{bounds}.
The NA49 lower bound on  $\Gamma[\Xi^{--}]$ is 18 MeV~\cite{Alt:2003vb}. Explanations for the unusually narrow widths vary.
In the chiral soliton model~\cite{Diakonov:1997mm} a cancellation between various coupling constants accounts for the
narrow width. This cancellation is argued to be exact in the large $N_c$ limit of QCD in
Ref.~\cite{Praszalowicz:2003tc}. In the quark model the width can be suppressed if the wavefunction of the pentaquark  is
such that the overlap with the ground state baryon and meson is small~\cite{Carlson:2003wc}.

Such cancellations are not manifest in the chiral Lagrangian and would appear as a fine tuning of the coupling
constant $g$. For pentaquarks with $J^\Pi = \frac{1}{2}^\pm$, $g^2 \sim 10^{-2}-10^{-3}$  is required to obtain
consistency with the most stringent bounds. For comparison, to reproduce the observed decay rates of the
ordinary decuplet baryons the  coupling constant is \mbox{$|g|\approx 2.1$}~\cite{Jenkins:1991ne}. For a
$J^\Pi=\frac{3}{2}^-$ pentaquark the two-body decay to the ground-state nucleon and meson could lie within
current experimental bounds without fine-tuning $g$. However, a $J^\Pi = \frac{3}{2}^-$ $\Xi^{--}$ pentaquark
can decay to $\Xi^-(1530) \pi^-$ via an $S$-wave. This decay
violates $SU(3)$ but not isospin. Though the smaller phase space for this decay and the $SU(3)$ suppression will
make this partial width significantly smaller than the $S$-wave partial width in Eq.~(\ref{rates}), these
factors are probably not enough to account for $\Gamma[\Xi^{--}] < 18$ MeV without an additional source of
suppression.

In the first part of Table I we give the partial widths for the two-body decays of  $\Xi^{--}$ and $\Xi^0$ normalized to
$\Gamma[\Theta^+]$.  The decays of the $\Xi^+$ , $\Xi^-$  are related  to $\Xi^{--}$, $\Xi^0$ decays by isospin factors:
\bea 
 \Gamma(\Xi^{+}_{\tenb} \ra \Xi^0 \pi^+)  & = & 
  \Gamma(\Xi^{--}_{\tenb} \ra \Xi^- \pi^-) \ , \\
 \Gamma(\Xi^{+}_{\tenb} \ra \Sigma^+ \ov{K}^0) &=& 
 \Gamma(\Xi^{--}_{\tenb} \ra \Sigma^- K^-)  \ , \nn\\
 \Gamma(\Xi^{-}_{\tenb} \ra \Xi^- \pi^0) = 
2 \; \Gamma(\Xi^{-}_{\tenb} \ra \Xi^0 \pi^-) &=& 
\Gamma(\Xi^{0}_{\tenb} \ra \Xi^0 \pi^0)   =  
  2 \; \Gamma(\Xi^{0}_{\tenb} \ra \Xi^- \pi^+) 
 \ ,  \nn \\
 \Gamma(\Xi^{-}_{\tenb} \ra \Sigma^0 K^-)  = 
2 \;  \Gamma(\Xi^{-}_{\tenb} \ra \Sigma^- \ov{K}^0) &=& 
  \Gamma(\Xi^{0}_{\tenb} \ra \Sigma^0 \ov{K}^0) =  
2 \; \Gamma(\Xi^{0}_{\tenb} \ra \Sigma^+ K^-) \ . \nn
\eea
\begin{table}[t]
\begin{tabular}{cccc}
\hline \hline
 & \multicolumn{3}{c}{$J^\Pi$} \\
\cline{2-4}
 & $ \qquad  \fr{1}{2}^- \qquad $ & $ \qquad  \fr{1}{2}^+ $  , \  $ \fr{3}{2}^+ \qquad$ & $  \fr{3}{2}^- $ \\[2mm]
\hline   & &  & \\[-2mm]
{\large $ \fr{\Gamma(\Xi^{--}_{\tenb} \ra \Xi^- \pi^-)}{\Gamma(\Theta^+)} $} 
 & $1.0 \pm 0.3$  & $4.0 \pm 1.2$ &$ 11. \pm 3.$ \\[4mm]
{\large $ \fr{\Gamma(\Xi^{--}_{\tenb} \ra \Sigma^- K^-)}{\Gamma(\Theta^+)} $} 
 &$ 0.84 \pm 0.25$ &$ 1.3 \pm 0.4 $ &$ 2.3 \pm 0.7$ \\[4mm]
{\large $ \fr{\Gamma(\Xi^{0}_{\tenb} \ra \Xi^- \pi^+)}{\Gamma(\Theta^+)} $} 
 &$ 0.33 \pm 0.10 $ &$ 1.3 \pm 0.4$ &$ 3.7 \pm 1.1$ \\[4mm]
{\large $ \fr{\Gamma(\Xi^{0}_{\tenb} \ra \Sigma^+ K^-)}{\Gamma(\Theta^+)} $} 
 & $0.29 \pm 0.09$  &$ 0.46 \pm 0.14$  & $0.88 \pm 0.26$ \\[4mm]
{\large $ \fr{\Gamma(\Xi^{0}_{\tenb} \ra \Xi^0 \pi^0)}{\Gamma(\Theta^+)} $} 
 & $0.68 \pm 0.20$  & $2.8 \pm 0.8$  & $ 7.9 \pm 2.4$  \\[4mm]
{\large $ \fr{\Gamma(\Xi^{0}_{\tenb} \ra \Sigma^0 \ov{K}^0)}{\Gamma(\Theta^+)} $} 
 & $0.56 \pm 0.17$ &$ 0.86 \pm 0.26$  &$ 1.6 \pm 0.5 $ \\[4mm]
\hline & & & \\[-2mm]
{\large $ \fr{\Gamma(\Xi^{--}_{\tenb} \ra \Xi^- \pi^-)}{\Gamma(\Xi^{--}_{\tenb} \ra \Sigma^- K^-)} $} 
 & $1.2 \pm 0.4$  & $ 3.1 \pm 0.9$ &  $4.7  \pm 1.4$ \\[4mm]
{\large $ \fr{\Gamma(\Xi^{0}_{\tenb} \ra \Xi^- \pi^+)}{\Gamma(\Xi^{0}_{\tenb} \ra \Sigma^+ K^-)} $} 
 & $1.1 \pm 0.3$ &$ 2.9 \pm 0.9$ & $4.2 \pm 1.3$  \\[4mm]
\hline& & &  \\[-2mm]
{\large $ \fr{\Gamma(\Xi^{--})}
             {\Gamma(\Theta^+ ) } $}
 & $> 1.8 \pm 0.5$   & $ >  5.3 \pm 1.6 $ &$ >  14. \pm 4.$  \\[4mm]
\hline
\hline
\end{tabular}
\caption{Partial widths and branching ratios for  $\Xi_{\tenb}^{--}, \Xi_{\tenb}^0$.} \label{table}
\end{table}
\noindent
We have not calculated  rates for the $N,\Sigma$ members of the $\tenb$ because there is no information about their masses
and because the analysis of these decays is complicated by the possibility  of these states mixing with nearby $N,\Sigma$
states from other $SU(3)$ multiplets. A nearly degenerate $\bf{8}$ of pentaquarks is predicted in the model of
Ref.~\cite{Jaffe:2003ci} and mixing with these states has been further studied in Ref.~\cite{Diakonov:2003jj}. The $\Xi_{\tenb}^-$
and $\Xi_{\tenb}^0$ will not mix with $\Xi_{\bf 8}^-$ and $\Xi_{\bf 8}^0$ because of isospin conservation. However, the
existence of such states will complicate the comparison of predictions for $\Xi_{\tenb}^-$ and $\Xi_{\tenb}^0$ decays  with
data if it is not possible  to separate $\Xi_{\tenb}^-$ and $\Xi_{\tenb}^0$ states from  $\Xi_{\bf 8}^-$ and
$\Xi_{\bf 8}^0$ states experimentally. If predictions for $\Xi^{--}$ and $\Xi^+$ decays work  but predictions for  $\Xi^-$
and $\Xi^0$ fail, this could be interpreted as a signal of a degenerate octet. Mixing with a nearly degenerate octet should
not affect $SU(3)$ relations between $\Theta^+$, $\Xi^{--}$ and $\Xi^+$ decays.  The errors quoted in Table~\ref{table} are
$\pm 30$\% which is the characteristic size of $SU(3)$ corrections. 

Calculation of two-body decays can be used to place the lower bounds on the ratio $\Gamma[\Xi]/\Gamma[\Theta^+]$ which are listed at
the bottom of Table~\ref{table}. In the case of $J^\Pi = \frac{3}{2}^-$ the  $SU(3)$ violating $S$-wave decay to
$\Xi(1530)$ could be comparable or even larger than the $D$-wave decay we have calculated, so the lower bound on
$\Gamma[\Xi]/\Gamma[\Theta^+]$ should be regarded  as especially weak. Similarly, for $J>\frac{3}{2}$ (except for $J^\Pi
=\frac{5}{2}^-$) the decay is through partial waves higher than $D$-waves and the suppression factor in Eq.~(\ref{sup})
gives an even larger lower bound on $\Gamma[\Xi]/\Gamma[\Theta^+]$. Therefore, if experiments eventually measure $\Gamma[\Xi]
< 10 \, \Gamma[\Theta^+]$, this will rule out $J^\Pi =\frac{3}{2}^-$ and $J > \frac{3}{2}$. The lower bounds for the ratio 
$\Gamma[\Xi]/\Gamma[\Theta^+]$ can potentially discriminate between $J^\Pi = \frac{1}{2}^+$ and $\frac{1}{2}^-$ as well.  

Also shown in Table~\ref{table} are ratios of partial widths of the $\Xi$ states. These are very interesting since they
clearly discriminate between  the $J^\Pi =\frac{1}{2}^+$ and $J^\Pi =\frac{1}{2}^-$ pentaquarks, which are the two
scenarios  considered most likely.   It is important to perform an experiment that can reconstruct both two-body decays of
the $\Xi$ so uncertainties associated with the production cross section cancel in the ratio allowing for an accurate
measurement of the branching fractions. The NA49 experiment observes $\Xi^{--} \to \Xi^- \pi^-$  but unfortunately can not
observe $\Xi^{--} \to \Sigma^- K^-$ because it lacks neutral particle detection to reconstruct the $\Sigma^-$ which decays to $n K^-$. 
A similar high energy experiment
with neutron detection capabilities may be able to perform such a measurement. Alternatively, in an experiment with photon
detection capabilities,  $\Xi^+$ could be reconstructed through the decays
\cite{price}:
\bea
\Xi^+ &\to & \Xi^0 \pi^+ \to \Lambda \pi^0 \pi^+ \to p \pi^- \gamma \gamma \pi^+ \ , \nn \\
&\to & \Sigma^+ \ov{K}^0 \to p \pi^0 \pi^+ \pi^- \to  p \gamma \gamma \pi^+ \pi^- \, .\nn
\eea
In photoproduction, one might hope to produce $\Xi^{--}$  through the processes $\gamma n \to \Xi^{--} K^+
K^+$~\cite{price}. In this experiment, the decay chain $\Xi^{--} \to \Sigma^- K^- \to n\pi^- K^-$ leaves only a
single neutral particle in the  final state, and since the process is exclusive the $n$ could be reconstructed
from the missing energy and momentum. We hope that this work motivates experimental measurement
of exotic $\Xi$ branching fractions. 

Finally, we mention other possible applications of the chiral Lagrangian introduced in this letter. One
possible application is computing nonanalytic chiral corrections to pentaquark masses. It would also be of
great interest to  make reliable predictions for pentaquark production cross sections. Unfortunately,
photoproduction of $\Theta^+$ occurs at such a high energy that the derivative expansion of chiral perturbation
theory is no longer a controlled expansion. The other important  production mechanism by which the $\Theta^+$
is observed  is the resonant reaction  $K^+ n \to \Theta^+ \to K^0 p$.  Adapting chiral perturbation theory
to resonant scattering is a straightforward application of the methods of Ref.~\cite{Bedaque:2003wa}. The
standard expressions for a resonant cross section will be recovered, along with nonresonant corrections coming
from contact interactions involving two kaons and two nucleons. The leading contact interaction comes from the
chiral covariant derivative in Eq.~(\ref{kin}); this gives a vanishing contribution to the $I=0$ channel where
the $\Theta^+$ appears. Therefore, higher dimension operators with two powers of $A_\mu$ and two $B$ fields
which are $O(Q^2)$ constitute the leading nonresonant contribution to $K N$ scattering. Since resonant
scattering occurs at $O(Q^{-1})$, the nonresonant contributions are   $O(Q^3/\Lambda_\chi^3)$ suppressed in
chiral perturbation theory. 

We would like to thank Roxanne Springer and Jos\' e Goity for useful comments on the manuscript.
C.S. and T.M. are supported in part by DOE grants
DE-FG02-96ER40945. T.M is also supported by DOE grant DE-AC05-84ER40150.

\end{document}